\documentstyle[12pt,aasms4]{article}
\newcommand{\lnls}{\hbox{Log N--Log S\ }}
\newcommand{\fun}{\hbox{\ erg cm$^{-2}$ s$^{-1}$} }
\newcommand{\lun}{\hbox{\ erg s$^{-1}$} }

\begin{document}

\title{The Chandra Deep Field South: the 1 Million Second 
  Exposure\altaffilmark{1}}
\altaffiltext{1}{Based on observations made at the European Southern
  Observatory, Paranal, Chile. The EIS observations have been carried
  out using the ESO New Technology Telescope (NTT) at the La Silla
  observatory (ESO LP 164.O-O561).
}

\author{P. Rosati\altaffilmark{1}, 
P. Tozzi\altaffilmark{2}, 
R. Giacconi\altaffilmark{3,4} 
R. Gilli\altaffilmark{3,5},
G. Hasinger\altaffilmark{6},
L. Kewley\altaffilmark{3},
V. Mainieri\altaffilmark{1,7},
M. Nonino\altaffilmark{2}, 
C. Norman\altaffilmark{3,8},
G. Szokoly\altaffilmark{6}, 
J.X. Wang\altaffilmark{3,9},
A. Zirm \altaffilmark{3}, 
J. Bergeron\altaffilmark{1},
S. Borgani\altaffilmark{10}, 
R. Gilmozzi\altaffilmark{1}, 
N. Grogin\altaffilmark{8}, 
A. Koekemoer\altaffilmark{8}, 
E. Schreier\altaffilmark{8}, and 
W. Zheng\altaffilmark{3} }

\affil{$^1$European Southern Observatory,
Karl-Schwarzschild-Strasse 2, D-85748 Garching, Germany}
\affil{$^2$Osservatorio Astronomico di Trieste, via G.B. Tiepolo 11,
I--34131, Trieste, Italy} 
\affil{$^3$Dept. of Physics and Astronomy, The Johns Hopkins
University, Baltimore, MD 21218, USA} 
\affil{$^4$ Associated
Universities Inc., 1400, 16th st. NW, Washington DC 20036, USA}
\affil{$^5$ Osservatorio Astrofisico di Arcetri, Largo E. Fermi 5,
I--50125 Firenze, Italy}
\affil{$^6$Astrophysikalisches Institut, An der Sternwarte 16, Potsdam
14482 Germany} 
\affil{$^7$ Dip. di Fisica, Universit\`a degli Studi Roma Tre, 
Via della Vasca Navale 84, I-00146 Roma, Italy}
\affil{$^8$ Space Telescope Science Institute, 3700
S. Martin Drive, Baltimore, MD 21210, USA} 
% \affil{$^7$ NRAO, 520 Edgemont Road, Charlottesville, Virginia}
\affil{$^9$ Center for Astrophysics, University of Science and 
Technology of China, Hefei, Anhui, 230026, P. R. China}
\affil{$^{10}$INFN, c/o
Dip. di Astronomia dell'Universit\`a, via Tiepolo 11, I--34131,
Trieste, Italy} 

\slugcomment{Accepted by {\it The Astrophysical Journal}, 16 October 2001}

\begin{abstract}

We present the main results from our 940 ksec observation of the
Chandra Deep Field South (CDFS), using the source catalog described in
an accompanying paper (Giacconi et al. 2001).  We extend the
measurement of source number counts to $5.5 \times 10^{-17}\fun$ in
the soft 0.5--2 keV band and $4.5 \times 10^{-16}\fun $ in the hard
2--10 keV band.  The hard band \lnls shows a significant flattening
(slope$\simeq 0.6$) below $\approx 10^{-14}\fun$, leaving at most
10--15\% of the X--ray background (XRB) to be resolved, the main
uncertainty lying in the measurement of the total flux of the XRB.  On
the other hand, the analysis in the very hard 5--10 keV band reveals a
relatively steep \lnls (slope $\simeq 1.3$) down to $10^{-15}\fun $.
Together with the evidence of a progressive flattening of the average
X--ray spectrum near the flux limit, this indicates that there is
still a non negligible population of faint hard sources to be
discovered at energies not well probed by {\it Chandra}, which
possibly contribute to the 30 keV bump in the spectrum of the XRB. We
use optical redshifts and identifications, obtained with the VLT, 
for one quarter of the sample
to characterize the combined optical and X--ray properties of the CDFS
sample.  Different source types are well separated in a parameter
space which includes X--ray luminosity, hardness ratio and $R-K$
color.  Type II objects, while redder on average than the field
population, have colors which are consistent with being hosted by a
range of galaxy types.  Type II AGN are mostly found at $z\lesssim 1$,
in contrast with predictions based on AGN population synthesis models,
thus suggesting a revision of their evolutionary parameters.

\end{abstract}


\section{Introduction}

With deeper and deeper exposures the {\it Chandra X--ray Observatory}
has resolved an increasingly large fraction of the X--ray background
(Mushotzky et al. 2000; Hornschemeier et al. 2000, 2001; Giacconi et
al. 2001--Paper I; Tozzi et al. 2001--Paper II; Brandt et al. 2001a),
particularly at high ($>2$ keV) energies. With the completion of two
1~Megasec {\it Chandra} exposures early this year, the Chandra Deep
Field South (CDFS, the subject of this Paper) and North (CDFN, Brandt
et al. 2001b), the original X--ray glow discovered by Giacconi et
al. in 1962 is now almost completely resolved in individual sources by
confusion-free {\it Chandra} images.  Observational and theoretical
studies over the last two decades have lead to a picture in which the
X--ray background (XRB) is due to accretion onto super-massive black
holes, integrated over cosmic time. With these new observations, the
general interest is now focusing on understanding the physical nature
and the cosmic evolution of these extragalactic sources, and their
role in models of galaxy evolution (e.g. Fabian 1999; Gilli, Salvati
\& Hasinger 2001).

In Paper I, we presented the first results from a 118 ks exposure. In
Paper II, we extended the statistical analysis of the X--ray
properties to 200 sources from a 300~ks exposure, and reported on the
optical identification and redshifts of a sizeable subsample from the
first results of our on--going spectroscopic program with the VLT. The
X--ray spectra were found to become progressively harder at fainter
fluxes, with an average stacked spectrum well fitted in the 1--10 keV
range by a power law with photon index $\Gamma=1.44\pm 0.03$, i.e.,
very close to the known spectral shape of the XRB (Marshall et
al. 1980; Gendreau et al. 1995).  The \lnls relation from the 300 ks
observations showed a significant flattening down to a flux limit of
$10^{-15} \fun$ when compared to previous ASCA determinations at
fluxes larger than $10^{-13} \fun$ (e.g. Cagnoni et al. 1998; Della
Ceca et al. 1999) and fluctuation analysis from Beppo SAX down to
$10^{-14} \fun$ (Perri \& Giommi 2000). These sources accounted for
60-90\% of the hard XRB, the main uncertainty lying in the measurement of
its absolute value with previous X--ray missions.  In Paper
II, we also found that the optical spectra and X--ray luminosities for
most of the hard sources are those typical of Type II AGN, in keeping
with expectations of AGN population synthesis models of the XRB
(e.g. Setti \&Woltjer 1989; Madau, Ghisellini \& Fabian 1993; Comastri
et al. 1995, Gilli, Salvati \& Hasinger 2001).

We present here the main results from the analysis of the 940 ks
exposure (hereafter 1~Msec) of the CDFS. In Section 2, the \lnls
relations are used to estimate the resolved fraction of the XRB in the
soft (0.5--2 keV) and hard (2--10 keV) bands.  In Section 3 we discuss
some of the optical properties of the source population in light of
the spectroscopic identification of 83 sources to date.  The
underlying catalog of X--ray sources used in this Paper is presented
in Giacconi et al. (2001, hereafter Paper III) along with optical
identifications in deep R band imaging data.

The spectroscopic dataset, which is being collected with on--going
observations with the FORS spectrograph at the VLT, is presented 
elsewhere (Szokoly et al. in prepration), while a detailed
analysis of the X--ray spectra of the CDFS sources and a comparison
with AGN models are presented in a forthcoming paper (Gilli et al. in
preparation).

\section{Statistical properties of the X--ray source population}

The Mega second dataset of the CDFS is the result of the coaddition of 11
individual {\it Chandra} ACIS--I (Garmire et al. 1992; Bautz et
al. 1998) exposures with aimpoints only a few arcsec from each other.
The nominal aim point of the CDFS is $\alpha=$3:32:28.0,
$\delta=-$27:48:30 (J2000).  We selected this field in a patch of the
southern sky characterized by a low galactic neutral hydrogen column
density ($N_H=8\times 10^{19}$ cm$^{-2}$) and a lack of bright
stars. The diary of observations is given in Paper III. The processing
method (data filtering, photon stacking, calibration procedures) is
described in Papers II and III. Due to different roll angles of the
individual pointings, the final image covers 0.109 deg$^2$.  The
decrease of effective exposure time and degradation of the angular
resolution at large off--axis angles cause the sky coverage to drop
below $10^{-15},\, 10^{-14} \fun$ in the soft and hard bands
respectively (Papers II and III).

In Figure \ref{fig1}, we show the color composite image of the CDFS.
This was constructed by combining images (first smoothed with a
Gaussian with $\sigma=1\arcsec$) in three bands (0.3--1 keV, 1--2 keV,
2--7 keV), which contain approximately equal numbers of photons from
detected sources.  Blue sources are those undetected in the soft
(0.5--2 keV) band, most likely due to intrinsic absorption from
neutral hydrogen with column densities $N_H > 10^{22}$ cm$^{-2}$. Very
soft sources appear red. A few extended low surface brightness sources
are also readily visible in the image.

Papers II and III describe in detail detection algorithms and
photometric procedures used.  By using a signal--to--noise threshold
of 2.1, we detected 346 sources in the 0.5--7 keV band; 307 have
$S/N>2.1$ in the soft band, 251 in the hard band and 110 in the very
hard (5--7 keV) band.  This S/N criterium was chosen to limit the
total number of spurious sources to less than 10, as established with
extensive simulations (see Paper II). The faintest detected sources
have approximately 10 counts in the soft band and 13 in the hard band.

For a power law source spectrum with a photon index $\Gamma = 1.4$,
absorbed by a Galactic column densities of $8\times 10^{19}
\hbox{cm}^{-2}$, we obtain a conversion factor of $(4.6\pm 0.1) \times
10^{-12} \fun$ per count s$^{-1}$ in the soft band (count rate in
0.5--2 keV to flux in the same band), $(3.0\pm 0.3) \times 10^{-11}
\fun$ per count s$^{-1}$ in the hard band (count rate in 2--7 keV to
flux in 2--10 keV), and $(1.02\pm 0.1) \times 10^{-10} \fun$ per count
s$^{-1}$ in the very hard band (count rate in 5--7 keV to flux in
5--10 keV). The assumed source spectrum is justified by the stacked
spectral analysis of all detected sources (see below).  The
uncertainty in the conversion factors corresponds to the observed
range of the average spectral index ($\Gamma = 1.4\pm 0.3$).

With these conversion factors, the on--axis flux limits for the soft,
the hard and the very hard bands are $5.5\times10^{-17}\fun $,
$4.5\times10^{-16} \fun$, and $10^{-15} \fun$ respectively.  These
values refer to the fluxes at which the sky coverage drops to zero for
the assumed detection threshold. We should stress that point sources
at fluxes fainter than these limits can statistically be detected in
the CDFS, at the expenses of increasing substantially the number of
spurious sources. This approach would not have any significant benefit
however, since the area covered at these fluxes is $\lesssim 5\%$ and
therefore only a handful of real sources would add to the sample. In
addition, the sky coverage becomes increasingly inaccurate at very
faint detection limits, making the faint end of the number counts very
uncertain.

In Paper III we also presented a catalog of sources characterized as
extended.  Visual inspection and optical colors indicate that some of
them are groups or isolated early type galaxies at $z<1$.  In several
cases however, the X--ray emission is dominated by a central, hard
component which is likely due to low--level nuclear activity.  The
mean surface brightness of these diffuse sources is as low as
$10^{-16}\fun \hbox{arcmin}^{-2}$, i.e.  between 10 and 20 times
fainter than the faintest extended sources discovered by ROSAT
(Rosati et al. 1995).

In Figure \ref{fig2}, we show the cumulative number counts in three
bands: soft, hard and very hard (5--10 keV).  The hard and soft
counts now cover several orders of magnitude in flux, therefore a
single power law is no longer a good fit. We perform a double power--law
fit to the differential number counts with four parameters: faint--end
normalization and slope, bright--end slope and flux where the break
occurs.  Likelihood contours for the faint end slope and normalization
are shown at the top of Figure \ref{fig2}.

In the soft band, the {\it Chandra} data extend the results from
previous surveys down to $5.5 \times 10^{-17} \fun$.  We find the
differential soft counts to be consistent with an euclidean slope at
the bright end, and a slope $\alpha_{diff} \equiv \alpha+1 = 1.63 \pm
0.13$ (1 $\sigma$ error) at the faint end, with a break at $S \simeq
1.3\times 10^{-14}\fun$.  Thus, below $S=10^{-15}$ erg s$^{-1}$
cm$^{-2}$, the slope of the cumulative number count is $\alpha \simeq
0.6$ (see top panels of Figure \ref{fig2}).  The normalization,
computed at $S=2\times 10^{-15}$ erg s$^{-1}$ cm$^{-2}$, is $K=380\pm
80$, which is consistent with previous values found in Papers I and
II.

The \lnls distribution for sources in the hard band is extended down
to a flux limit of $4.5 \times 10^{-16}$ erg s$^{-1}$ cm$^{-2}$ with
the present data.  As in Paper II, the hard counts are normalized at
the bright end ($S> 10^{-13} \fun$) with the ASCA data by Della Ceca
et al. (1999).  The best fit bright--end slope is consistent with an
Euclidean slope, whereas the faint end has a slope of $\alpha_{diff} =
1.61 \pm 0.10$.  The break occurs at $S\simeq 8\times 10^{-15}\fun$
and $K=1300\pm 100$.  This confirms a further flattening when compared
with the 300 ks CDFS exposure (Paper II) and shallower {\it Chandra}
fields (e.g. Stern et al. 2001a; Mushotzky et al. 2000; see the top
panels in Figure \ref{fig2}).
 
In order to test the impact of cosmic variance in these relatively
small fields ($\simeq 0.1 \, \hbox{deg}^2$), we compare our faint
number counts with those derived from the public CDFN data using our
method of analysis (i.e. photon file processing, detection algorithm,
etc.).  The slope and normalization of the faint end are found to be
consistent within two sigma level (see dashed contours in the top
panels of Figure \ref{fig2}), whereas the CDFN counts at bright fluxes
are significantly higher than in the CDFS. We find $\alpha=0.53\pm
0.1$ and $K=490\pm 90$ at the faint end of the soft counts, and
$\alpha=0.61\pm 0.10$ and $K=1500\pm 100$ at the faint end of the hard
counts.  Our best fit values for the CDFN \lnls are in very good
agreement with those recently published by Brandt et al. (2001b) with
a completely different method of detection, calibration and sky
coverage computation. This suggests that differences in number counts
are not due to systematics in methods of analysis, but rather to field
to field variations.

We compared our \lnls with the predictions of the AGN population
synthesis models described in Gilli, Salvati \& Hasinger (2001).  We
consider their model B, where the number ratio $R$ between absorbed
and unabsorbed AGNs is assumed to increase with redshift from 4, the
value measured in the local Universe, to 10 at $z\sim 1.3$, where $R$
is unknown.  This model was found to better describe observations with
respect to a standard model where $R=4$ at all redshifts.  The value
of R is assumed to be independent of the luminosity, and therefore a
large population of obscured QSOs is included in such a model.  In
this Paper, the parameters of the AGN X--ray luminosity functions
assumed in the model B are renormalized to fit the background
intensity measured by HEAO--1 ($1.6 \fun\rm{deg}^{-2}$ in the 2--10
keV band).  The predictions for the source counts (solid line in
Figure \ref{fig2}) in the hard and soft bands are in very good
agreement with the {\it Chandra} \lnls at all fluxes.

The resolved fraction of the hard X--ray background, after including
the bright end as observed by ASCA (Della Ceca et al. 1999), amounts
to $(1.57 \pm 0.15) \times 10^{-11}$ erg s$^{-1}$ cm$^{-2}$
deg$^{-2}$.  This value, which is computed down to the limiting flux of
$S=4.5\times 10^{-16}$ erg s$^{-1}$ cm$^{-2}$, has now reached, if not
exceeded, the value of the total (unresolved) X--ray background $1.6
\times 10^{-11}$ erg s$^{-1}$ cm$^{-2}$ deg$^{-2}$ measured by UHURU
and HEAO--1 (Marshall et al. 1980).  If we assume more recent
measurements of the integrated background from ASCA surveys (Ishisaki
et al. 1999; Ueda et al. 1999), the resolved fraction of the hard
XRB is 80--90\%. The BeppoSAX measurement (Vecchi et al. 1999; see
Paper II) is likely too high to be accounted for by our number counts.

Field to field variations have also a non negligible impact on these
measurements. For example, we have shown that the differences at the
bright end, and to a lesser extent at the faint end, of the \lnls in
the CDFS and CDFN fields are likely an effect of cosmic variance.  The
total contribution to the XRB in the CDFN amounts to $(1.7 \pm
0.17)\times 10^{-11}$ erg s$^{-1}$ cm$^{-2}$ deg$^{-2}$, which is 8\%
higher than the CDFS value.

The existence of a fainter and progressively harder population of
sources which are likely to fill the remaining fraction of the hard
XRB is confirmed by the steep counts in the very hard band (5--10
keV), also plotted in Figure \ref{fig2}.  A single power--law fit
yields a slope of $\alpha = 1.35 \pm 0.15$, with a normalization
$K=940\pm 100$, somewhat lower than the value obtained with XMM
observations of the Lockman Hole in the same band (Hasinger et
al. 2001), and lower than the extrapolation of the bright 5--10 keV
counts of the Hellas2XMM survey (Baldi et al. 2001).  The best fit
values in the CDFN are $\alpha = 1.18 \pm 0.20$ and $K=1230\pm 200$.
A double power law fit indicates a break at $S\simeq 4\times 10^{-15}$
erg s$^{-1}$ cm$^{-2}$, but with large uncertainties on the slopes due
to the smaller number of sources (110) and to the smaller flux range
probed.  The steeper slope with respect to the 2-10 keV number counts
is mainly due to the hardening of the average spectrum at low fluxes.
In fact, by running our detection algorithm in the 5--7 keV band
alone, we find only 5 additional sources which are just below the
detection threshold in the 2--7 or 0.5--7 keV bands.  Forthcoming XMM
observations of the CDFS will be crucial to investigate the nature of
these sources which, in light of our first spectroscopic
identifications, are expected to be strongly absorbed AGN, at
relatively low redshift

It is also interesting to quote the contribution of all CDFS sources
to the soft band, which we restrict to 1--2 keV to be compared with
the measurement of the total flux measured by ROSAT, $4.4\times
10^{-12}\fun\rm{deg}^{-2}$ (see Hasinger et al. 1998).  We find a
resolved contribution of $\simeq 6.25 \times 10^{-13}\fun$ deg$^{-2}$
for fluxes lower than $10^{-15}$ erg s$^{-1}$ cm$^{-2}$, corresponding
to $\simeq 14$\% of the total value.  When we include the integrated
contribution of point--like sources at higher fluxes as measured by
ROSAT, the total resolved contribution at fluxes larger than $3\times
10^{-17}\fun $ (our flux limit in the 1--2 keV band) adds up to
$\simeq 3.65 \times 10^{-12}\fun$deg$^{-2}$, corresponding to 83\% of
the total value.  In addition, X--ray clusters are known to contribute
6\% of the ROSAT 1--2 keV background (10\% in the 0.5--2 keV band, see
Rosati et al. 1998).  This sets an upper limit of $\simeq 11$\% for
the diffuse emission from warm gas.  Incidentally, we note that we
resolved completely the value found by ASCA (Gendreau et al. 1995),
which, however, must be considered a lower limit to the 1--2 keV total
background (see discussion in Hasinger et al. 1998).

\section{Spectral properties, X--ray and optical colors}

The cumulative spectral properties of the CDFS sample show an
increasing hardness at fainter fluxes (Paper I and II; Brandt et al
2001b).  The analysis of the stacked spectrum of the total sample (see
Paper II for technical details) gives now an average photon index of
$\Gamma = 1.375 \pm 0.015$ (error at 90\% confidence level) with
$\chi^2_{\nu}=1.65$ in the 1--10 keV range, after fixing the absorbing
column density to the Galactic value ($N_H = 8\times 10^{19}$
cm$^{-2}$).  The average spectrum is consistent with the shape of the
total hard background $\langle \Gamma \rangle \simeq 1.4$, confirming
previous findings.  To quantify the progressive hardening of the
spectral shape as a function of the hard flux, we divided the sample
of sources which are detected in the hard band in 4 subsamples,
defined in order to have roughly the same number of net source counts
($\simeq$ 3000): bright ($S> 2\times 10^{-14}$ erg s$^{-1}$
cm$^{-2}$), medium ($ 2\times 10^{-14}>S> 6\times 10^{-15}$ erg
s$^{-1}$ cm$^{-2}$), faint ($ S< 6\times 10^{-15}$ erg s$^{-1}$
cm$^{-2}$), and very faint ($ S< 2\times 10^{-16}$ erg s$^{-1}$
cm$^{-2}$, with only $\simeq 1500$ counts).  The best fit slope of the
stacked spectra is $\Gamma = 1.68\pm 0.03$, $1.42\pm 0.03$, $1.10\pm
0.03$, and $1.05\pm 0.05$ respectively.  Incidentally, by assuming a
varying $\Gamma$ as a function of the flux, we verify that the changes
in the \lnls are well within the quoted errors (less than 5\%).

In Figure \ref{fig3} we show the hardness ratio as a function of the
luminosity in the 0.5--10 keV band for 83 sources for which we have
optical spectra and secure classification (Szokoly et al. in
preparation).  The hardness ratio is defined as $HR \equiv
(H-S)/(H+S)$ where $H$ and $S$ are the net count rates in the hard
(2--7 keV) and the soft band (0.5--2 keV), respectively.  The X--ray
luminosities are not corrected for internal absorption and are
computed in a critical universe with $H_0=50$ km s$^{-1}$ Mpc$^{-1}$.
Different source types are clearly segregated in this plane.  Type I
AGNs (marked with circles) have luminosities between $10^{43}$ erg
s$^{-1}$ and $10^{45}$ erg s$^{-1}$, with hardness ratios in a narrow
range around $HR\simeq -0.5$.  This corresponds to an effective
$\Gamma = 1.7$, commonly found in Type I AGN.  Type II AGN are skewed
towards significantly higher hardness ratios ($HR\gtrsim 0$), with
(absorbed) luminosities in the range $10^{42}-10^{44}$ \lun.  Note
that the unabsorbed, intrinsic luminosities of Type II AGN would fall
in the same range as that of Type I's (see Gilli et al. in preparation).
A subset of objects characterized by spectra of normal galaxies
(marked with triangles) have very low luminosities ($\simeq 10^{41}$
erg s$^{-1}$) and very soft spectra (several with $HR=-1$), 
as expected in the case of starbursts or thermal halos. 
However, a separate subset has harder spectra ($HR > -0.5$),
and luminosities larger than $10^{42}$ erg s$^{-1}$.  In these
galaxies the X--ray emission is likely due to a mixture of low level
AGN activity and a population of low mass X--ray binaries (see also
Fiore et al. 2000; Barger et al. 2001).  In Figure \ref{fig3}, we also
include two Type II QSOs (asterisks), which are the most luminous
sources in the HR--L$_X$ plane with HR$>0$. We discovered one in the
CDFS itself (Norman et al. 2001); the second one was recently found by
Stern et al. (2001b) in the Lynx field.

In Figure \ref{fig4}, we present a color-magnitude diagram, $K$ vs
$R-K$, for all CDFS sources with available photometry and optical
counterparts. Vega magnitudes are derived from our survey with
FORS--VLT (Program-ID 64.O-0621) and from the ESO Imaging Survey (EIS,
Vandame et al. 2001; Arnouts et al. 2001).  Optical spectral
classification is used as in Figure~\ref{fig3}. As a control sample,
field galaxies are also shown. Type~II sources are, on average,
significantly redder than Type~I sources and the overall field
population, as expected in objects with dust and gas obscured AGN
activity (e.g. Alexander et al. 2001).  As a reference, on the same
diagram we plot evolutionary tracks of an early and three late type
galaxies using the template library of Coleman, Wu, Weedman (1980)
(for E/S0, Sbc, Scd and Irr), whose spectral energy distributions
(SEDs) have been extended to the near-IR and far UV using Bruzual
\&Charlot (2000) models.  Magnitudes are normalized to the measured
local value $K^\ast=10.8$, no dust extinction is assumed.  The QSO
track was obtained using an empirical template from the Sloan Digital
Sky Survey (Vanden Berk et al. 2001).  The QSO SED was extended in the
near IR using the Granato et al. (1997) models and normalized to
$M_B^\ast=-22.4$.

The color-magnitude region spanned by the evolutionary tracks of
$L_\ast$ galaxies ranging from late to early types encompasses most of
the CDFS sources (with the exception of the bluest Type I objects
which lie along the QSO track). Similarly, the template tracks in the
$B-R$ vs $R-K$ plot in Figure \ref{fig5} cover the color space of the
optical counterparts of the CDFS source. This coverage becomes even
more uniform if a range of dust extinctions is taken into
account. This suggests that Type I and Type II AGN are hosted by a
broad range of galaxy types.  These findings are in agreement with the
results of Schreier et al. (2001) and of Koekemoer et al. (2001) who
found {\sl morphological types} of CDFS sources to be uncorrelated
with their hardness ratio from high-resolution HST imaging data.

\section{Conclusions}

In this Paper, we have reported the main results from the 1 Ms
exposure of the Chandra Deep Field South, using the catalog of 346
sources presented by Giacconi et al. (2001).  These observations reach
a flux limit of $5.5 \times 10^{-17}\fun $ in the 0.5--2 keV band,
$4.5\times 10^{-16}$ erg s$^{-1}$ cm$^{-2}$ in the 2--10 keV band (or
$3.6 \times 10^{-16} \fun$ in the 2--8 keV band), and $10^{-15}\fun $
in the 5--10 keV band.

The source number counts in the hard (2--10 keV) band show a
significant flattening ($\alpha\simeq 0.6$) below $S \simeq 8\times
10^{-15}\fun$.  The two deepest {\it Chandra} surveys, CDF South and
North (Brandt et al. 2001b), have substantially resolved the hard XRB
in individual sources, the main uncertainty now lying in the
measurement of its total value.  Nonetheless, we find the number
counts at the hardest energies probed by {\it Chandra} (5--10 keV
band) to be significantly steep ($\alpha\simeq 1.3$). Together with
the evidence that the progressive steepening of the average X--ray
spectrum of the faint population seems to continue at fluxes near or
below our flux limit, this indicates that there is still a not
negligible population of faint hard sources to be discovered with
better sensitivity at high energies. Such sources are expected to
generate the 30 keV bump in the XRB spectrum. XMM may play an
important role in unveiling this population, if not affected by
confusion.
%A large number of Chandra studies have convincingly shown
%that $\sim\!1\arcsec$ resolution is an essential key element in deep
%surveys.

We have shown that fluctuations in the number counts among different
{\it Chandra} fields can be as large as $\sim 5\%$ at low fluxes and
up to $\sim 30\%$ at bright fluxes (a more detailed analysis of this
effect is presented in Tozzi et al. 2001b).  These differences
correspond to 1--2 $\sigma$ Poissonian fluctuations (see top panels of
Figure~\ref{fig2}).  Fluctuations of counts in excess of the Poisson
noise are expected as a result of significant source clustering over
scales comparable to the field size.  By extrapolating the detected
angular two--point correlation function for $X$--ray sources (e.g.,
Vikhlinin et al. 1995; Giacconi et al. 2001) out to the typical scale
of {\it Chandra} fields, we find the contribution of clustering
to cosmic variance to be negligible.  

Using optical identification and redshifts for a quarter of the total
sample, our on--going spectroscopic observations indicate that the
bulk of the obscured Type II AGN (with HR$>0$) lies at a median
redshift $\approx 0.8$ (Paper II; see also Crawford et al. 2000;
Barger et al. 2001).  A stacked spectrum analysis of these sources
reveals that they are characterized by an intrinsic absorption with
$\log N_H >21.5$, assuming a median redshift of $z= 1$.  These
sources are on average redder than the field galaxy population,
however, they still span a portion in color--color diagrams which is
covered by a broad range of host galaxy spectral types. The fact that
Type II AGN may not be segregated just within one morphological class
of galaxies is corroborated by HST/WFPC2 imaging of several of these
sources (Koekemoer et al. 2001).

Even accounting for the incomplete spectroscopic identification of our
sample, the redshift ditribution of sources is in stark contrast with
AGN population synthesis models which predicts a peak at redshifts
$1.3 - 2$ (e.g. Gilli, Salvati \& Hasinger 2001).  A similar
distribution at low redshift has been observed in the Lockman Hole
(Hasinger et al. 2001).  Together with the evidence that the observed
effective spectral slope of the faint X--ray population is harder than
the model predictions (see Figure 8 in Paper II), this suggests that
such models will have to be modified to incorporate different
evolutionary scenarios (Gilli et al. in preparation).

A substantial fraction (11\%) of the CDFS sources remain unidentified
even in deep VLT optical images ($R\lesssim 27.5$) or moderately deep
near--IR imaging (a 15\% fraction down to $K\simeq 22$).  The hardness ratio
distribution of these ``blank'' sources is similar to the overall
distribution in the CDFS, making it difficult to comment on the nature
of these sources with current data.

A great deal of spectroscopic work with 8--10m class telescopes, as
well as deep multi-wavelength photometry to constrain redshifts and
SEDs of the faintest population ($\sim\! 30\%$ of the CDFS sources are
fainter than $R=25$), will be needed to characterize the
nature and the evolution of the AGN population. Scheduled mid--IR
observations with SIRTF of the two Chandra Deep fields (the GOODS
project, Dickinson et al. 2001) will play a prominent role in
this study, by probing heavily obscured objects at virtually any
redshift.

\acknowledgements

We thank Dr. Harvey Tananbaum and the entire {\it Chandra } X--ray
Center Team for the high degree of support we have received in
carrying out our observing program.  In particular, we wish to thank
Antonella Fruscione for her constant help in the use of the CXC
software.  R. Giacconi and C. Norman gratefully acknowledge support
under NASA grant NAG--8--1527 and NAG--8--1133.

\newpage

\begin{figure}
% \plotone{xcol_940k.ps}
\caption{A color composite image of the Chandra Deep Field South of
940 ks (pixel size=0.984\arcsec, smoothed with a $\sigma=1\arcsec$
Gaussian).  The image was obtained combining three energy bands:
0.3--1 keV, 1--2 keV, 2--7 keV (respectively red, green and blue).  A
few diffuse reddish (i.e. soft) sources, associated with groups of
galaxies can be seen.  The color intensity is derived directly from
the net counts and has not been corrected for vignetting.
\label{fig1}}
\end{figure}

\clearpage

\begin{figure}
\centerline{\epsfxsize=.7\textwidth \epsfbox{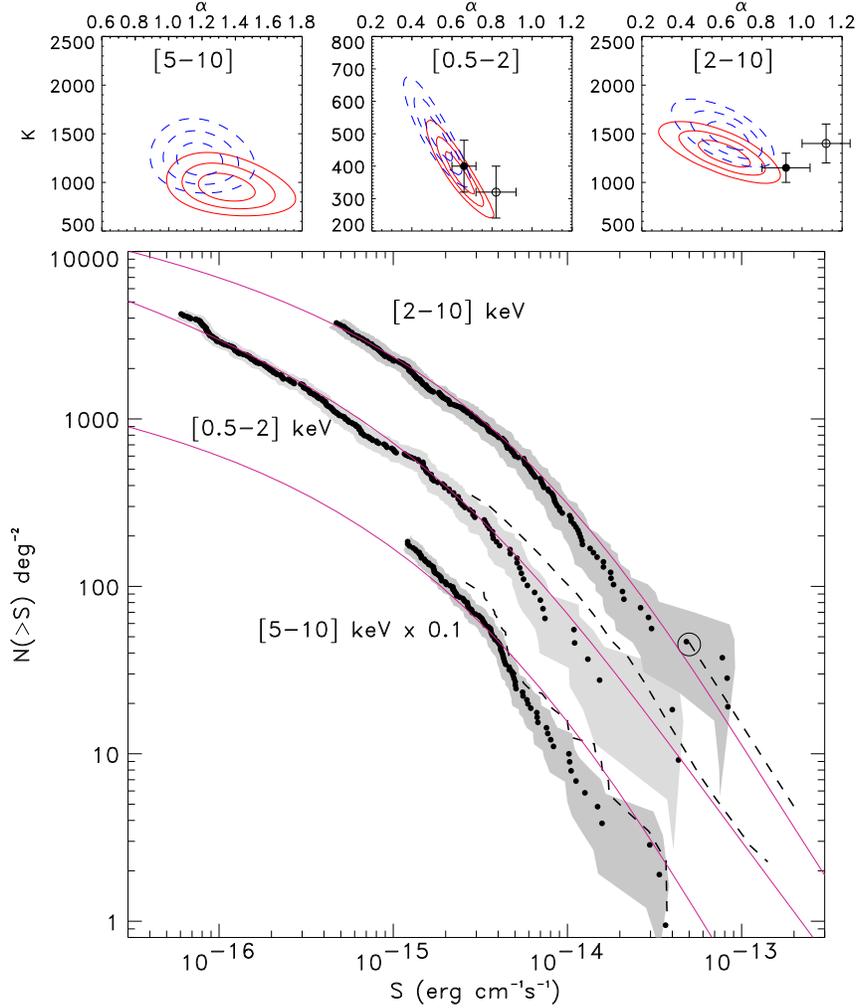}}
\caption{\protect\small \lnls in the soft (0.5--2 keV), hard (2--10
keV) and very hard (5--10 keV) bands from the 1 Megasec observations
of the CDFS.  The shaded areas indicate uncertainties due to the
Poisson statistics (1 $\sigma$) and the choice of the conversion
factor (corresponding to a power law spectrum with $\Gamma = 1.4\pm
0.3$). Dashed lines are previous measurements: ROSAT deep surveys in
the 0.5--2 keV band (Hasinger et al. 1998), ASCA surveys in the
2--10 keV band (Cagnoni et al. 1998; Della Ceca et al. 1999) and the
recent XMM Lockman Hole observation in the 5--10 keV band (Hasinger
et al. 2000).  Solid lines are predictions from model B by Gilli,
Salvati \& Hasinger (2001) (see text for details).  The three top
panels show the likelihood contours for the best fit normalization and
slope of the faint end in the three bands from the CDFS (solid) and
CDFN (dashed) data. Data points refer to the best fit values of Paper
II (300 ks, filled circles) and the Lynx field (190 ks, Stern et
al. 2001, open circles). }
\label{fig2}
\end{figure}

\clearpage

\begin{figure}
\plotone{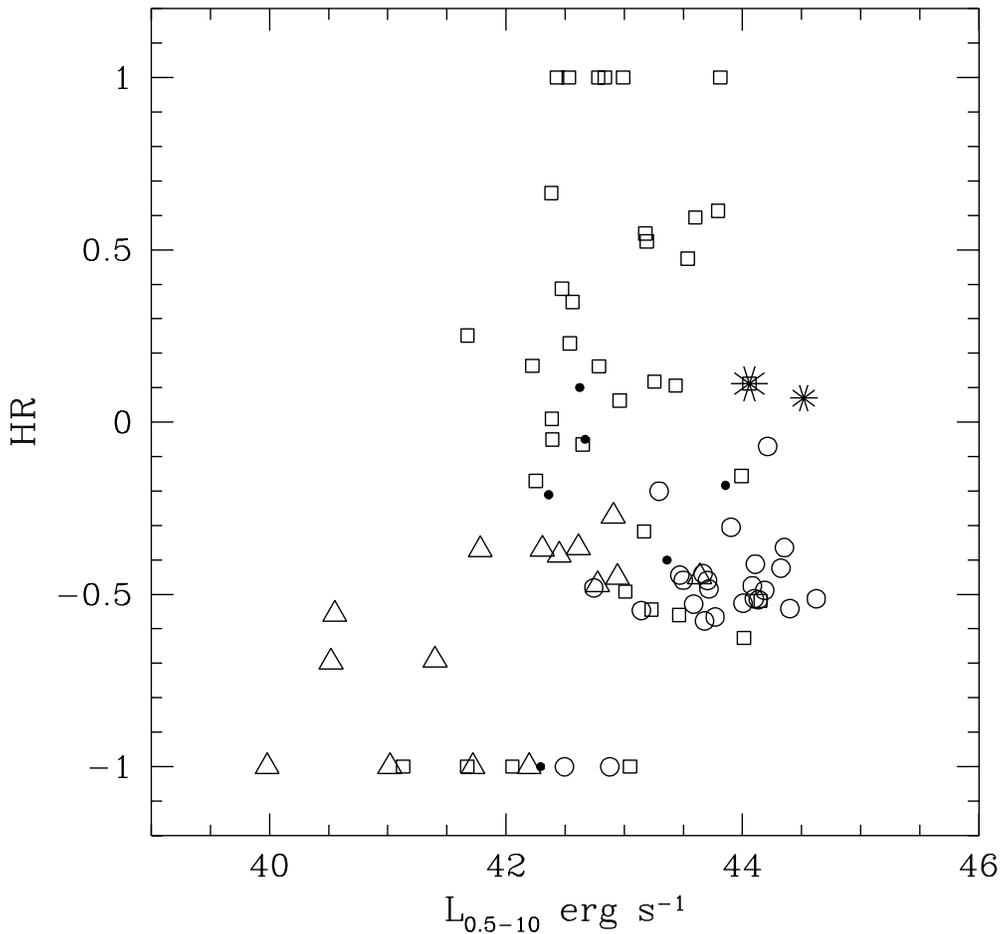}
\caption{Hardness ratio versus rest frame luminosity in the total
0.5--10 keV band.  Type I AGNs are marked with circles, Type II with
squares, and ``normal galaxies'' with triangles.  Unidentified sources
are marked with dots.  The large and small asterisks indicate the two
Type II QSOs found in the CDFS (Norman et al. 2001) and the Lynx field
(Stern et al. 2001), respectively.  A critical density universe with
$H_0 = 50$ km s$^{-1}$ Mpc$^{-1}$ has been adopted.
Luminosities are not corrected for possible intrinsic absorption.  }
\label{fig3}
\end{figure}

\clearpage

\begin{figure}
\plotone{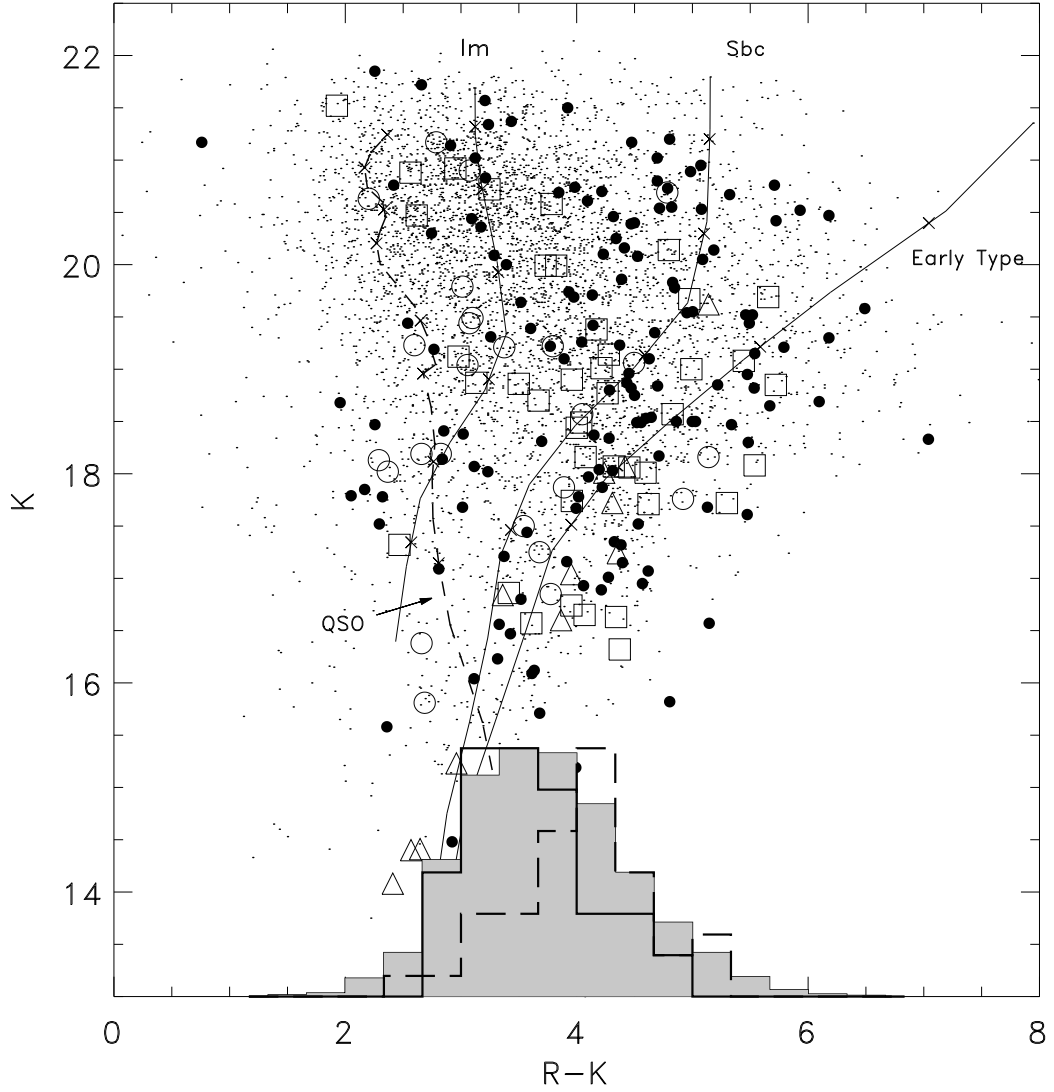}
\caption{Color-magnitude diagram, $K$ vs $R-K$, of CDFS sources.
Magnitudes are in the Vega system.  Symbols for {\it Chandra} sources are as
in Figure 3, while field sources are marked with small dots.  The
histogram shows the color distribution of the field sources (shaded
area) compared with the distribution of Type I AGNs (continuous line)
and Type II AGNs (dashed line).  The three evolutionary tracks
correspond to an unreddened QSO with $L=L^\ast_B$ ($z=0\div 5$), and
unreddened late type and early type $L^\ast$ galaxies ($z=0\div 2.8$)
from the Coleman, Wu, Weedman (1980) template library. Cross
signs along the tracks mark the redshifts $0.5, 1.0, 1.5, \ldots$ . }
\label{fig4}
\end{figure}

\clearpage

\begin{figure}
\plotone{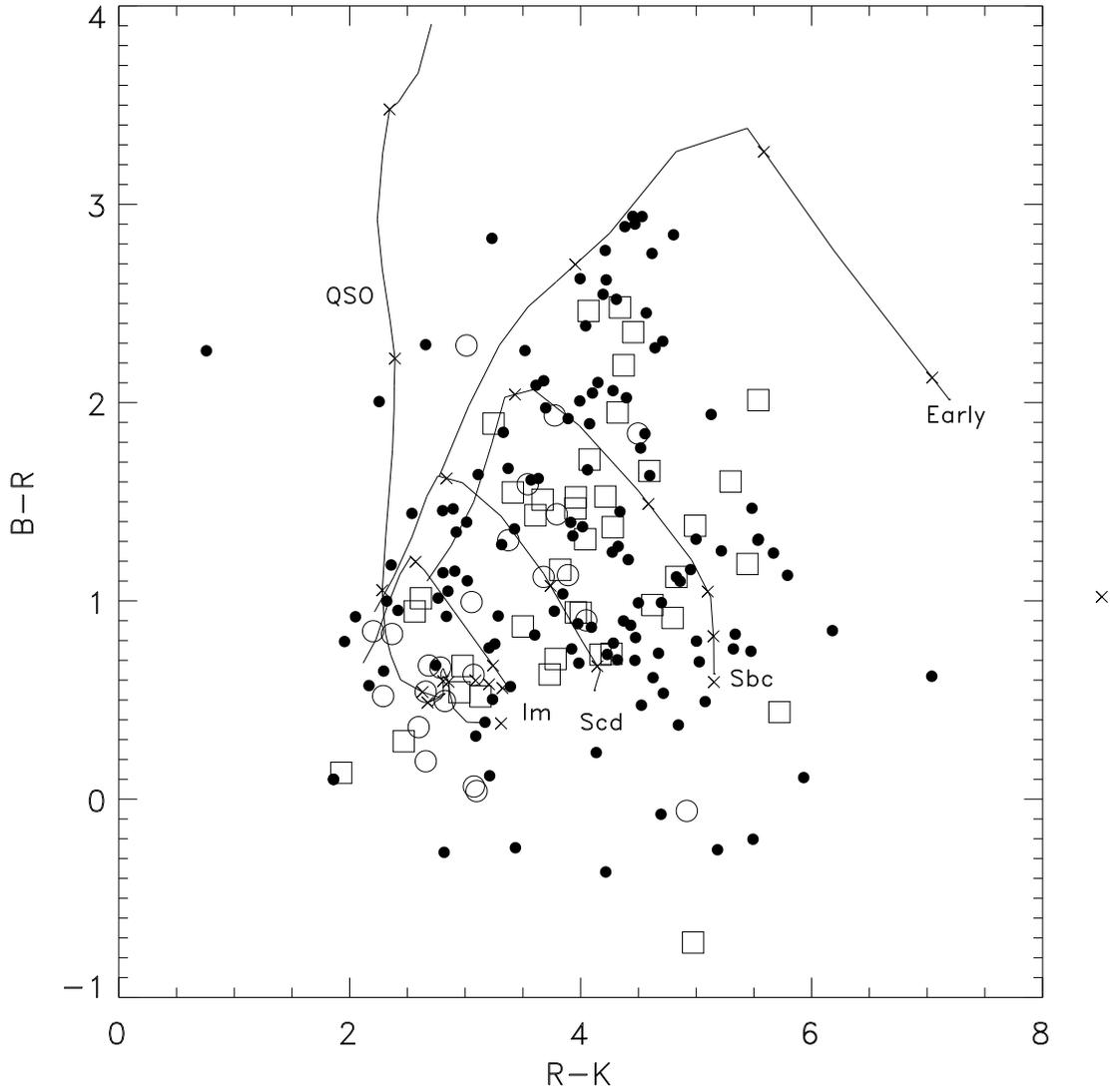}
\caption{Color-color diagram, $B-R$ vs $R-K$, of CDFS sources.
Symbols for {\it Chandra} sources are as in Figure~3; evolutionary tracks of
E/S0, Sbc, Scd, Irr galaxies are as in Figure~4.  }
\label{fig5}
\end{figure}

\end{document}